\documentclass[conference,10pt]{IEEEtran}
\usepackage{cite}
\usepackage{epsfig}
\usepackage{epstopdf}
\usepackage{subfig}
\usepackage{graphicx}
\usepackage{xcolor}
\usepackage{tikz}
\usetikzlibrary{arrows,positioning,shapes.geometric,circuits.logic.US}
\usepackage{pgfplots}
\usepackage{multirow}
\usepackage{upgreek}
\usepackage{amssymb}
\usepackage{amsmath}
\usepackage{mathabx}
\usepackage{cases}
\usepackage{url}
\usepackage{pifont}
\usepackage{bm}
\usepackage{amsthm}

\usepackage{tabularx}
\usepackage{booktabs}
\usepackage{filecontents}
\newcolumntype{Y}{>{\centering\arraybackslash}X}

\usepackage{algorithm} 
\usepackage[noend]{algpseudocode}
\usepackage{array}

\newcommand{\fixme}[2]{\ifx&#2&{\leavevmode\color{red}#1}\else{\leavevmode\color{red}FIXME\{}#1{\leavevmode\color{red}\}}\footnote{{\leavevmode\color{red}#2}}\PackageWarning{Fixme}{#1: #2}\fi}

\newcommand{\newstuff}[2]{\ifx&#2&{\leavevmode\color{blue}#1}\else{\leavevmode\color{blue}FIXME\{}#1{\leavevmode\color{blue}\}}\footnote{{\leavevmode\color{blue}#2}}\PackageWarning{Newstuff}{#1: #2}\fi}

\DeclareMathOperator*{\sgn}{sgn}

\title{Sliding Window Polar Codes}
\author{\IEEEauthorblockN{ Valerio~Bioglio, Carlo~Condo, Ingmar Land\\}
\IEEEauthorblockA{Mathematical and Algorithmic Sciences Lab\\ Huawei Technologies France SASU \\
Email: $\{$valerio.bioglio,carlo.condo, ingmar.land$\}$@huawei.com}} 
\begin{document}

\maketitle
\thispagestyle{empty}
\begin{abstract}
We propose a novel coupling technique for polar codes via a special kernel that enables efficient sliding window decoding.
This feature allows to reduce the memory requirement of the decoder, an important possibility in wireless communication downlink scenarios. 
Our approach is based on the design of an ad-hoc kernel to be inserted in a multi-kernel polar code framework. 
Simulation results show that the proposed sliding window polar codes 
outperform polar codes transmitted in independent blocks at a negligible additional decoding overhead. 
\end{abstract}

\IEEEpeerreviewmaketitle

\section{Introduction} \label{sec:intro}
Polar codes \cite{arikan} are linear block codes relying on the channel polarization effect, a phenomenon that creates virtual bit-channels of different reliability. 
In a polar code of length $N$ and dimension $K$, the information is transmitted through the $K$ most reliable bit-channels, while the remaining bit-channels are ``frozen" to a predetermined value, usually zero. 
Originally, polar codes were based on the polarization kernel $T_2 = \begin{bmatrix}
   1 & 0 \\
   1 & 1
\end{bmatrix}$, which limited code length $N$ to be a power of 2. 
This limitation has been overcome thanks to the discovery of polarization properties of larger kernels, for which kernels providing the best polarization properties have been identified \cite{kernel2}.
In parallel, multi-kernel polar codes \cite{mk_conf,multi-kernel-journal} have been proposed to create polar codes of virtually any length  with efficient implementations and fast decoding \cite{MK_dec,MK_fast}. 

In this paper, we propose a novel design for polar codes to make them decodable through a sliding window framework, i.e. considering $M<N$ received symbols per decoding step, and using the result of one step to facilitate the next.
The proposed method may have practical application in a wireless communication downlink scenario, where the receiver has usually less computational capabilities than the transmitter. 
Recent works have already shown the benefits of inserting memory in the polar code encoding and decoding process \cite{PCmemory,PCmemory2}. 
Moreover, windowed decoding has been proposed for the decoding of spatially coupled codes \cite{win_spat} and for convolutional LDPC codes \cite{tc_LDPC}; a similar approach has been proposed in \cite{pic_polar} to construct partially information coupled polar codes.

Our approach differs from the aforementioned techniques, as it is based on the definition of an ad-hoc kernel enabling the sliding window decoding of a particular multi-kernel polar code. 
The analysis of the polarization effect of large kernels permits to construct very efficient polar codes \cite{large_ker_faz} with low decoding complexity \cite{eff_dec_large_ker,large_ker_win_dec}.
Even if it is possible to perform this analysis on the kernel presented in this paper, the exact evaluation of its polarization exponent \cite{scal_exp_large_ker} is not fundamental for the sliding window nature of the proposed method, so it is postponed to a further paper. 
Our novel contributions are: (i) introduction of a kernel allowing for sliding window decoding; (ii) code design for this construction; (iii) discussion of encoding and decoding; (iv) numerical analysis. 
As we show through both theoretical analysis and simulations, the proposed code design is able to outperform transmission with independent blocks at negligible decoding overhead.

\section{Background} \label{sec:prel}

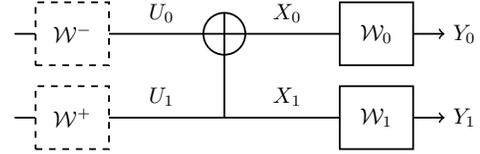
\begin{figure}[tb]
	\begin{center}
		\resizebox{0.35\textwidth}{!}{\begin{tikzpicture}[scale=0.33, line width=0.8pt]

	\draw[->] (-7,8) -- (13.5,8) node[right] {$Y_0$};
	\draw[->] (-7,4) -- (13.5,4) node[right] {$Y_1$};
	
	\draw[fill=white,dashed]   (-6,9.5) rectangle (-2.5,6.5) node[midway]{$\mathcal{W}^-$};
	\draw[fill=white,dashed]   (-6,5.5) rectangle (-2.5,2.5) node[midway]{$\mathcal{W}^+$};
	
	\draw[fill=white]   (8.5,9.5) rectangle (12,6.5) node[midway]{$\mathcal{W}_0$};
	\draw[fill=white]   (8.5,5.5) rectangle (12,2.5) node[midway]{$\mathcal{W}_1$};
	
	\draw[] (3,4)                     -- (3,9) ;
	\draw  (3,8) ellipse (1 and 1);

\node at (0,9) {$U_0$};
\node at (0,5) {$U_1$};
\node at (6,9) {$X_0$};
\node at (6,5) {$X_1$};
\end{tikzpicture}}
	\end{center}
	\caption{Basic combination of two IB-DMC channels.}
	\label{fig:basic_trans}
\end{figure}
The combination of two independent binary discrete memoryless channels (IB-DMCs) $\mathcal{W}_0$ and $\mathcal{W}_1$ defined over alphabet $\mathcal{X} = \{0,1\}$, $\mathcal{W}_i : \mathcal{X} \rightarrow \mathcal{Y}$, depicted in Figure~\ref{fig:basic_trans}, is performed by the channel transformation matrix $T_2$. 
Transition probabilities for the split channels $\mathcal{W}^-$ and $\mathcal{W}^+$ are 
\begin{eqnarray}
\mathcal{W}^- (y_0,y_1|u_0) = \frac{1}{2} \sum_{u_1 \in \mathcal{X}} \mathcal{W}_0 (y_0|u_0 \oplus u_1) \mathcal{W}_1 (y_1|u_1)~,\\
\mathcal{W}^+ (y_0,y_1,u_0|u_1) = \mathcal{W}_0 (y_0|u_0 \oplus u_1) \mathcal{W}_1 (y_1|u_1)~.
\end{eqnarray}
The one-step transformation defined by $T_2$ can be rewritten using operators $\boxcoasterisk$ and $\oasterisk$ introduced in \cite{korada_thesis} to degrade and enhance channel parameters as
\begin{eqnarray}
\mathcal{W}^- = \mathcal{W}_0 \boxcoasterisk \mathcal{W}_1 ~,\\
\mathcal{W}^+ = \mathcal{W}_0 \oasterisk \mathcal{W}_1 ~.
\end{eqnarray}
The Bhattacharyya parameter of channel $\mathcal{W}_i$ is defined as
\begin{equation}
Z(\mathcal{W}_i) = \sum_{y \in \mathcal{Y}} \sqrt{\mathcal{W}_i(y|0)\mathcal{W}_i(y|1)}~
\end{equation} 
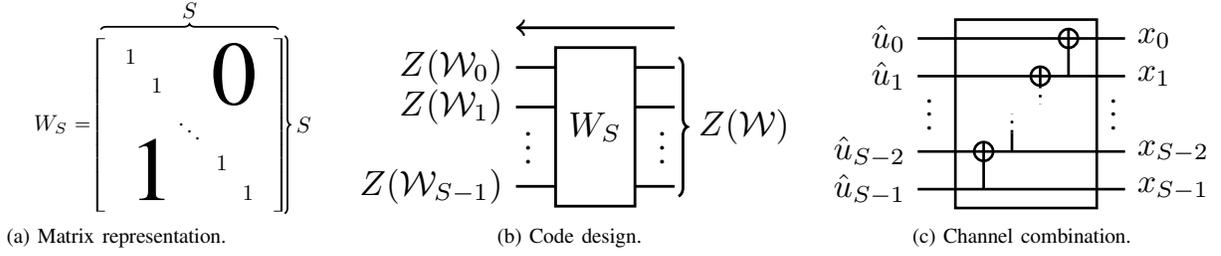
\begin{figure*}[t]
	\subfloat[Matrix representation.]
		{\label{fig:mat} \resizebox{0.3\textwidth}{!}{\usetikzlibrary{decorations.pathreplacing,angles,quotes}
\usetikzlibrary{matrix, decorations.pathreplacing}
\begin{tikzpicture}[scale=0.2, line width=0.8pt]
	\matrix (vec) [matrix of math nodes, left delimiter = {[}, right delimiter = {]}] {
	        1   \\
 	           & 1    \\ 
 	           &    &  \ddots    \\
 	           &    &             & 1   \\
	           &    &             &    & 1 \\
	};
	
 	\draw [decorate,decoration={brace,mirror}] (7.5,-7) -- (7.5,7);
 	\node at (0,9) {$S$};
 	\draw [decorate,decoration={brace}] (-7,7.5) -- (7,7.5);
 	\node at (-10,0) {$W_S=$};
 	\node at (9,0) {$S$};
	\node[font=\Huge,scale=1.7] at (3.5,4) {0};
	\node[font=\Huge,scale=1.7] at (-3,-3.5) {1};
\end{tikzpicture}}}
	\subfloat[Code design.]
		{\label{fig:des_block} \resizebox{0.35\textwidth}{!}{\usetikzlibrary{decorations.pathreplacing,angles,quotes}
\begin{tikzpicture}[scale=0.2, line width=0.8pt]
	\draw[<-] (-2,9)                     -- (6,9) ;
	\draw[]  (0,0) rectangle (4,8) node[midway]{$W_S$};
	\draw[] (-2,7) node[left] {$Z(\mathcal{W}_0)$} -- (0,7) ;
	\draw[] (-2,5) node[left] {$Z(\mathcal{W}_1)$} -- (0,5) ;
	\draw[] (-0.5,3.5) node[left] {$\vdots$}    (4.5,3.5) node[right] {$\vdots$} ;
	\draw[] (-2,1) node[left] {$Z(\mathcal{W}_{S-1})$} -- (0,1) ;
	\draw[] (4,7)                     -- (6,7) ;
	\draw[] (4,5)                     -- (6,5) ;
	\draw[] (4,1)                     -- (6,1) ;
	\node at (9.5,4) {$Z(\mathcal{W})$};
	\draw [decorate,decoration={brace,mirror}] (6.2,0.5) -- (6.2,7.5);
\end{tikzpicture}}}
	\subfloat[Channel combination.]
		{\label{fig:graph} \resizebox{0.3\textwidth}{!}{\begin{tikzpicture}[scale=0.2, line width=0.8pt]
	\draw   (0,-2) rectangle (7.5,8) ;
	\draw[] (-2,7) node[left] {$\hat{u}_0$} -- (9,7) node[right] {$x_0$};
	\draw[] (-2,5) node[left] {$\hat{u}_1$} -- (9,5) node[right] {$x_1$};
	\draw[] (-0.5,3.5) node[left] {$\vdots$}  (7.5,3.5) node[right] {$\vdots$}  ;
	\draw[] (-2,1) node[left] {$\hat{u}_{S-2}$} -- (9,1) node[right] {$x_{S-2}$};
	\draw[] (-2,-1) node[left] {$\hat{u}_{S-1}$} -- (9,-1) node[right] {$x_{S-1}$};
	\draw[] (6,5)                     -- (6,7.5) ;
	\draw  (6,7) ellipse (0.5 and 0.5);
	\draw[dotted] (4.5,4.5)                     -- (4.5,3.5) ;
	\draw[] (4.5,4.5)                     -- (4.5,5.5) ;
	\draw  (4.5,5) ellipse (0.5 and 0.5);
	\draw[] (3,1)                     -- (3,2) ;
	\draw[dotted] (3,2)                     -- (3,3) ;
	\draw[] (1.5,-1)                     -- (1.5,1.5) ;
	\draw  (1.5,1) ellipse (0.5 and 0.5);
\end{tikzpicture}}}	
	\caption{Kernel $W_S$ description.}
	\label{fig:block-W}
\end{figure*}
and can be used to evaluate the channel reliability, since it provides an upper bound on the probability that an error occurs under ML decoding \cite{arikan}. 
The Bhattacharyya parameter of the transformed channels can be calculated as 
\begin{align}
\label{eq:oast}
Z(\mathcal{W}_0 \boxcoasterisk \mathcal{W}_1) & \leq 1 - \left( 1 - Z(\mathcal{W}_0) \right) \cdot \left( 1 - Z(\mathcal{W}_1) \right)~, \\
Z(\mathcal{W}_0 \oasterisk \mathcal{W}_1) &= Z(\mathcal{W}_0) Z(\mathcal{W}_1)~.
\end{align}

Polar codes rely on the polarization acceleration enabled by the $n$-fold Kronecker product of the basic channel transformation kernel $T_2$; for a polar code length $N=2^n$, its channel transformation matrix is given by $T_N = T_2^{\otimes n}$. 
As the code length goes toward infinity, bit-channels become completely noisy or completely noiseless, and the fraction of noiseless bit-channels approaches the channel capacity. 
In case of finite code lengths, however, the polarization of bit-channels is incomplete, generating bit-channels that are partially noisy. 
The Bhattacharyya parameter of these intermediary bit-channels can be tracked throughout the polarization stages to estimate their reliabilities. 
For an $(N,K)$ polar code, the indices of the $K$ most reliable bit-channels are selected to form the information set $\mathcal{I}$. 
The input vector $u = [u_0,u_1,\ldots,u_{N-1}]$ is then created by assigning the $K$ message bits to the entries of $u$ whose indices are listed in $\mathcal{I}$; the remaining entries of $u$, forming the frozen set $\mathcal{F}$, are set to zero. 
The codeword $x$ is finally calculated as $x = u \cdot T_N$. 

Polar codes can be decoded through Successive Cancellation (SC) \cite{arikan}, that can be described as a depth-first binary tree search, with priority given to the left branch. 
To improve the error-correction performance of SC at moderate code lengths, SC list (SCL) decoding has been proposed in \cite{tal_list}.
It relies on $L$ parallel SC decoders working on different paths. 
Every time an information bit is estimated, the paths are doubled, with $L$ decoders considering it a $0$, and $L$ a $1$.
Based on path metrics, only the $L$ best paths are kept.  After the last information bit has been decoded, the path with the best metric is output.

\section{Sliding Window Polar Codes} \label{sec:SWPC}
\begin{figure*}[tb]
	\begin{center}
	    \resizebox{0.85\textwidth}{!}{\usetikzlibrary{decorations.pathreplacing,angles,quotes}
\begin{tikzpicture}[scale=0.2, line width=0.8pt]
	\draw [decorate,decoration={brace,mirror}] (-9.5,7.5) -- (-9.5,-1);
	\node at (-11.5,3.5) {$u^{(1)}$};
	\draw [decorate,decoration={brace,mirror}] (-9.5,-1.5) -- (-9.5,-10);
	\node at (-11.5,-5.5) {$u^{(2)}$};
	\draw [decorate,decoration={brace,mirror}] (-9.5,-15.5) -- (-9.5,-24);
	\node at (-11.5,-19.5) {$u^{(S)}$};
	\draw[] (-2,7) node[left] {$u_0$} -- (6,7) ;
	\draw[] (-2,4.5) node[left] {$u_1$} -- (6,4.5) ;
	\draw[] (-0.5,3) node[left] {$\vdots$}    (4.5,3) node[right] {$\vdots$} ;
	\draw[] (-2,0.5) node[left] {$u_{M-1}$} -- (6,0.5) ;
	\draw[fill=white]  (0,-0.5) rectangle (4,8) node[midway]{$T_M$};
	
	\draw[] (-2,-2) node[left] {$u_M$} -- (6,-2) ;
	\draw[] (-2,-4.5) node[left] {$u_{M+1}$} -- (6,-4.5) ;
	\draw[] (-0.5,-6) node[left] {$\vdots$}    (4.5,-6) node[right] {$\vdots$} ;
	\draw[] (-2,-8.5) node[left] {$u_{2M-1}$} -- (6,-8.5) ;
	\draw[fill=white]  (0,-9.5) rectangle (4,-1) node[midway]{$T_M$};
	
	\draw[] (3,-12.5) node[left] {$\vdots$}     ;
	
	\draw[] (-2,-16.5) node[left] {$u_{N-M}$} -- (6,-16.5) ;
	\draw[] (-2,-19) node[left] {$u_{N-M+1}$} -- (6,-19) ;
	\draw[] (-0.5,-20.5) node[left] {$\vdots$}    (4.5,-20.5) node[right] {$\vdots$} ;
	\draw[] (-2,-23) node[left] {$u_{N-1}$} -- (6,-23) ;
	\draw[fill=white]  (0,-24) rectangle (4,-15.5) node[midway]{$T_M$};
	
	\draw[] (13,7) node (v2) {} -- (21,7) ;
	\draw[] (13,4.5) -- (21,4.5) ;
	\draw[] (14.5,3) node[left] {$\vdots$}    (19.5,3) node[right] {$\vdots$} ;
	\draw[] (13,0.5) -- (21,0.5) ;
	\draw[fill=white]  (15,-0.5) rectangle (19,8) node[midway]{$W_S$};
	
	\draw[] (13,-2) -- (21,-2) ;
	\draw[] (13,-4.5) -- (21,-4.5) ;
	\draw[] (14.5,-6) node[left] {$\vdots$}    (19.5,-6) node[right] {$\vdots$} ;
	\draw[] (13,-8.5) -- (21,-8.5) ;
	\draw[fill=white]  (15,-9.5) rectangle (19,-1) node[midway]{$W_S$};
	
	\draw[] (18,-12.5) node[left] {$\vdots$}     ;
	
	\draw[] (13,-16.5) -- (21,-16.5) ;
	\draw[] (13,-19) -- (21,-19) ;
	\draw[] (14.5,-20.5) node[left] {$\vdots$}    (19.5,-20.5) node[right] {$\vdots$} ;
	\draw[] (13,-23) -- (21,-23) ;
	\draw[fill=white]  (15,-24) rectangle (19,-15.5) node[midway]{$W_S$};
	
	\draw[] (6,7) -- (13,7) ;
	\draw[] (6,4.5) -- (13,-2) ;
	\draw[] (6,0.5) -- (13,-16.5) ;
	\draw[] (6,-2) -- (13,4.5) ;
	\draw[] (6,-4.5) -- (13,-4.5) ;
	\draw[] (6,-8.5) -- (13,-19) ;
	\draw[] (6,-16.5) -- (13,0.5) ;
	\draw[] (6,-19) -- (13,-8.5) ;
	\draw[] (6,-23) -- (13,-23) ;
	
	\draw[] (21,7) -- (28,7) ;
	\draw[] (21,4.5) -- (28,-2) ;
	\draw[] (21,0.5) -- (28,-16.5) ;
	\draw[] (21,-2) -- (28,4.5) ;
	\draw[] (21,-4.5) -- (28,-4.5) ;
	\draw[] (21,-8.5) -- (28,-19) ;
	\draw[] (21,-16.5) -- (28,0.5) ;
	\draw[] (21,-19) -- (28,-8.5) ;
	\draw[] (21,-23) -- (28,-23) ;
	
	\draw[] (28,7) -- (30,7) node[right] {$x_{0}$};
	\draw[] (28,4.5) -- (30,4.5) node[right] {$x_{1}$};
	\draw[] (28,0.5) -- (30,0.5) node[right] {$x_{M-1}$};
	\draw[] (28,-2) -- (30,-2) node[right] {$x_{M}$};
	\draw[] (28,-4.5) -- (30,-4.5) node[right] {$x_{M+1}$};
	\draw[] (28,-8.5) -- (30,-8.5) node[right] {$x_{2M-1}$};
	\draw[] (28,-16.5) -- (30,-16.5) node[right] {$x_{N-M}$};
	\draw[] (28,-19) -- (30,-19) node[right] {$x_{N-M+1}$};
	\draw[] (28,-23) -- (30,-23) node[right] {$x_{N-1}$};
	
	\draw[] (30.5,3) node[left] {$\vdots$};
	\draw[] (30.5,-6) node[left] {$\vdots$};
	\draw[] (33.5,-12.5) node[left] {$\vdots$};
	\draw[] (30.5,-20.5) node[left] {$\vdots$};


	\draw[] (32.5,7) -- (42.5,7) node[right] {$y_{0}$};
	\draw[] (32.5,4.5) -- (42.5,4.5) node[right] {$y_{1}$};
	\draw[] (35,0.5) -- (42.5,0.5) node[right] {$y_{M-1}$};
	\draw[] (33.5,-2) -- (42.5,-2) node[right] {$y_{M}$};
	\draw[] (35,-4.5) -- (42.5,-4.5) node[right] {$y_{M+1}$};
	\draw[] (35.5,-8.5) -- (42.5,-8.5) node[right] {$y_{2M-1}$};
	\draw[] (35.5,-16.5) -- (42.5,-16.5) node[right] {$y_{N-M}$};
	\draw[] (37.5,-19) -- (42.5,-19) node[right] {$y_{N-M+1}$};
	\draw[] (35,-23) -- (42.5,-23) node[right] {$y_{N-1}$};
	
      \draw[fill=white] (38.5,8) rectangle (41,6)  node[midway]{$\mathcal{W}$};
      \draw[fill=white] (38.5,5.5) rectangle (41,3.5)  node[midway]{$\mathcal{W}$};	
      \draw[fill=white] (38.5,1.5) rectangle (41,-0.5)  node[midway]{$\mathcal{W}$};
      \draw[fill=white] (38.5,-1) rectangle (41,-3)  node[midway]{$\mathcal{W}$};	
      \draw[fill=white] (38.5,-3.5) rectangle (41,-5.5)  node[midway]{$\mathcal{W}$};
      \draw[fill=white] (38.5,-7.5) rectangle (41,-9.5)  node[midway]{$\mathcal{W}$};	
      \draw[fill=white] (38.5,-15.5) rectangle (41,-17.5)  node[midway]{$\mathcal{W}$};	
      \draw[fill=white] (38.5,-18) rectangle (41,-20)  node[midway]{$\mathcal{W}$};
      \draw[fill=white] (38.5,-22) rectangle (41,-24)  node[midway]{$\mathcal{W}$};		

	\draw [decorate,decoration={brace,mirror}] (50,-1) -- (50,7.5);
	\node at (52.5,3.5) {$y^{(1)}$};
	\draw [decorate,decoration={brace,mirror}] (50,-10) -- (50,-1.5);
	\node at (52.5,-5.5) {$y^{(2)}$};
	\draw [decorate,decoration={brace,mirror}] (50,-24) -- (50,-15.5);
	\node at (52.5,-19.5) {$y^{(S)}$};
	
	\draw [dashed] (5.25,3.75) ellipse (0.5 and 4.25);
	\node at (6.75,8.5) {$t^{(1)}$};
	\draw [dashed] (5.25,-5.25) ellipse (0.5 and 4.25);
	\node at (5.75,-10.5) {$t^{(2)}$};
	\draw [dashed] (5.25,-19.75) ellipse (0.5 and 4.25);
	\node at (6.75,-24.5) {$t^{(S)}$};
	
\end{tikzpicture}}
	\end{center}
	\caption{Tanner graph of transformation matrix $T$ of sliding window polar code.} 
	\label{fig:trans_mat}
\end{figure*}
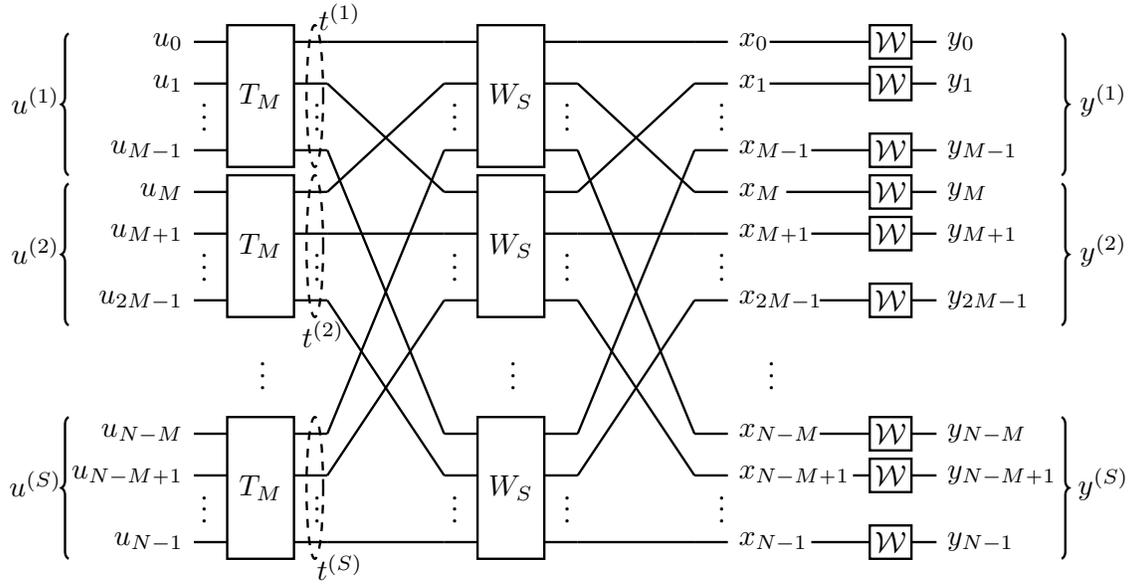

\subsection{Sliding window kernel} \label{sec:W_S}

The sliding window kernel $W_S$ is the cornerstone of the proposed construction. 
It is defined by the full binary lower triangular matrix of size $S$, i.e. the square matrix of size $S \times S$ having ones on and below the diagonal, and zeros above the diagonal, as depicted in Figure~\ref{fig:mat}. 
The matrix $W_S$ imposes a channel transformation that can be described as a cascade of basic channel transformations, as represented in Figure~\ref{fig:graph}. 
This structure imposes to decode input bit $u_i$ using only two channel likelihoods out of $S$, namely $\tilde{x}_i$ and $\tilde{x}_{i+1}$, the former being modified by the hard decisions taken on bits $0\le j<i$. 
We will see how to leverage on this property to construct a multi-kernel polar code mixing $W_S$ with classical polar kernel $T_2^{\otimes n}$ to enable for a sliding window decoding mechanism. 

The virtual channel $\mathcal{W}_i$ experienced by input bit $u_i$ transmitted through channel transformation matrix $W_S$ over channel $\mathcal{W}$ undergoes the transformation
\begin{equation}
\mathcal{W}_i = \mathcal{W} \boxcoasterisk \underbrace{(\mathcal{W} \oasterisk \mathcal{W} \oasterisk \ldots \oasterisk \mathcal{W})}_{i+1 \text{ terms}}, 
\end{equation}
for $0 \leq i < S-1$, while for the last channel
\begin{equation}
\mathcal{W}_{S-1} = \underbrace{\mathcal{W} \oasterisk \mathcal{W} \oasterisk \ldots \oasterisk \mathcal{W}}_{S \text{ terms}}.  
\end{equation}
The Bhattacharyya parameter of these virtual channels, depicted in Figure~\ref{fig:des_block}, can hence be calculated as
\begin{equation}
Z(\mathcal{W}_{i-1}) \leq
  \begin{cases}
    1 - (1 - Z(\mathcal{W})) \cdot (1 - Z(\mathcal{W})^{i}) & \text{if } i < S,\\
    Z(\mathcal{W})^S & \text{if } i = S.
  \end{cases} \label{eq:W}
\end{equation}

The Bhattacharyya parameter can be used to evaluate the bit-channel reliability for various channel models. 
For binary erasure channels (BECs), the bit error probabilities can be directly calculated, while under additive white Gaussian noise (AWGN) channels, density evolution under Gaussian approximation (DE/GA) technique can be implemented on their basis \cite{trip_DEGA}; this method estimates the likelihood distribution of the polarized channels by tracking their mean at each stage of the SC decoding tree. 
Given the block representation of kernel $W_S$ depicted in Figure \ref{fig:graph}, the bit erasure probability $\delta_i$ of bit $u_i$ under BEC transmission can be calculated as 
\begin{equation}
\delta_{i-1} =
  \begin{cases}
    1 - (1 - \delta) \cdot (1 - \delta^i) & \text{if } i < S~,\\
    \delta^S & \text{if } i = S~,
  \end{cases} \label{eq:delta}
\end{equation}
where $\delta$ is the bit erasure probability of the original BEC. 
For the AWGN channel, the equations tracking the log-likelihood mean $\mu_{i}$ of input bit $u_i$ for $W_S$ are given by
\begin{equation}
\mu_{i-1} =
  \begin{cases}
    \phi^{-1} \left( 1 - (1 - \phi(\mu)) \cdot (1 - \phi(i \mu)) \right) & \text{if } i < S, \\
    S \mu & \text{if } i = S,
  \end{cases} \label{eq:mu}
\end{equation}
where $\phi(\cdot)$ is defined as
\begin{equation}
\small
  \phi(x) = 
    \begin{cases}
    1-\frac{1}{2\sqrt{\pi x}} \int_{-\infty}^{\infty} \tanh \left( \frac{z}{2} e^{\frac{(z-x)^2}{4x}} \right) dz & \text{if } x > 0,\\
	1                             & \text{if } x = 0,
    \end{cases}
\end{equation}
and can be approximated as described in \cite{comp_AWGN}. 

\subsection{Code design} \label{sec:des}

As for classical polar codes, the design of a sliding window polar code entails the selection of its transformation matrix $T$ and frozen set $\mathcal{F}$. 
Given $S=N/M$, the transformation matrix of the code is defined as $T=W_S \otimes T_M$, where $T_M=T_2^{\otimes m}$ is the transformation matrix of a polar code of length $M$. 
A sliding window polar code can hence be described as a particular multi-kernel polar code \cite{mk_conf} constructed by placing kernel $W_S$ before the transformation matrix of a classical polar code of length $M$. 
Note that the position of $W_S$ with respect to the other kernels is not interchangeable. 
The Tanner graph of transformation matrix $T$ is depicted in Figure~\ref{fig:trans_mat}. 

The frozen set can be designed according to the general multi-kernel polar code approach \cite{mk_conf} using equations described in Section~\ref{sec:W_S}.
Given the structure of $T$, however, the calculation of channel bit reliabilities can be simplified as follows.  
The Tanner graph depicted in Figure~\ref{fig:trans_mat} shows that the channel transformation imposed by $T$ can be seen as the juxtaposition of $S$ polar code transformations of length $M$, each one altering a different channel whose reliability depends on $W_S$. 
Given the transmission channel $\mathcal{W}$, it is thus possible to initially calculate the Bhattacharyya parameter of the virtual channel $\mathcal{W}_s$ experienced by the $s$-th polar code $\mathcal{P}_s$ defined by $T_M$ using \eqref{eq:W}; then, classical polar codes design equations can be used to evaluate the bit-channel reliabilities of input bits $u_{(s-1)M},u_{(s-1)M+1},\ldots,u_{sM-1}$. 
With reference to Figure \ref{fig:trans_mat}, the Battacharyya parameters at the left of each $T_M$ block would be independently calculated using the inputs at their right, that have already undergone one further polarization step through $W_S$.  
Finally, all the bit-channels of input vector $u$ are sorted in order of reliability, where the indices of the $N-K$ least reliable bit-channels form the frozen set $\mathcal{F}$, and the indices of the remaining $K$ bit-channels form the information set $\mathcal{I}$ of the code.

\subsection{Encoding} \label{sec:enc}
Encoding of the proposed sliding window polar codes can be performed as for standard polar codes.
The $K$ message bits are inserted in the input vector $u$ according to the information set previously calculated, namely storing their values in the indices listed in $\mathcal{I}$, while the remaining bits of $u$ are set to zero. 
The codeword $x$ can be calculated as $x = u \cdot T$.

However, the particular structure of $W_S$ allows for an alternative encoding algorithm based on previously described set of $S$ polar codes $\mathcal{P}_1,\dots,\mathcal{P}_S$ of size $M$. 
The information set $\mathcal{I}_s$ of polar code $\mathcal{P}_s$ can be extracted from the global information set $\mathcal{I}$ as the set of entries of $\mathcal{I}$ comprised between $(s-1)\cdot M$ and $s \cdot M - 1$, 
\begin{equation}
	\mathcal{I}_s = \{ i - (s-1) \cdot M \, | \, i \in I, (s-1) \cdot M \leq i < s \cdot M \} ,
	\label{eq:Is}
\end{equation}
for $s = 1,\dots,S$. 
Similarly, partial input vectors $u^{(s)}$ 
are extracted from $u$ or can be created on the basis of the message bits. 
Each partial input vector can be encoded independently through polar encoding of length $M$, obtaining $S$ partial codewords $t^{(1)},\dots,t^{(S)}$. 
Finally, the codeword $x$ is obtained by backward accumulation of the partial codewords, starting from the last one, i.e. $x=[t^{(1)}\oplus \ldots \oplus t^{(S)},t^{(2)}\oplus \ldots \oplus t^{(S)},\ldots,t^{(S-1)} \oplus t^{(S)},t^{(S)}]$.

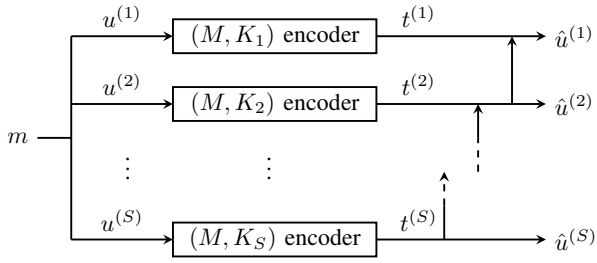
\begin{figure}[tb]
	\begin{center}
	    \resizebox{0.45\textwidth}{!}{\begin{tikzpicture}

\draw[thick] (-1,-0.5) node[left] {$m$} -- (-0.5,-0.5);
\draw[thick] (-0.5,1) -- (-0.5,-2);

\draw[->, >=stealth, thick] (-0.5,1) -- (1,1) node[midway, above]{$u^{(1)}$};
\draw[->, >=stealth, thick]  (-0.5,0) -- (1,0) node[midway, above]{$u^{(2)}$};
\draw[->, >=stealth, thick] (-0.5,-2) -- (1,-2) node[midway, above]{$u^{(S)}$};
\draw[] (0.5,-0.85) node[left] {$\vdots$}    (2.25,-0.85) node[right] {$\vdots$} ;

\draw[thick] (1,1.25) rectangle (4,0.75) node[midway]{$(M,K_1)$ encoder};
\draw[thick] (1,0.25) rectangle (4,-0.25) node[midway]{$(M,K_2)$ encoder};
\draw[thick] (1,-1.75) rectangle (4,-2.25) node[midway]{$(M,K_S)$ encoder};

\draw[->, >=stealth, thick]  (4,1) -- (6.5,1) node[right] {$\hat{u}^{(1)}$} node[near start, above]{$t^{(1)}$};
\draw[->, >=stealth, thick]  (6,0) -- (6,1);
\draw[->, >=stealth, thick]  (4,0) -- (6.5,0) node[right] {$\hat{u}^{(2)}$} node[near start, above]{$t^{(2)}$};
\draw[->, >=stealth, thick]  (5.5,-0.5) -- (5.5,0);
\draw[thick, dashed]  (5.5,-1) -- (5.5,-0.5);
\draw[->, >=stealth, thick] (4,-2) -- (6.5,-2) node[right] {$\hat{u}^{(S)}$} node[near start, above]{$t^{(S)}$};
\draw[->, >=stealth, thick, dashed]  (5,-1.5) -- (5,-1);
\draw[thick]  (5,-2) -- (5,-1.5);


\end{tikzpicture}}
	\end{center}
	\caption{Encoding scheme of sliding window polar codes.}
	\label{fig:enc}
\end{figure}

This encoding strategy, depicted in Figure~\ref{fig:enc}, follows the structure of the sliding-window kernel $W_S$ and permits to maintain the classical polar encoding structure, however slightly increasing the encoding latency. 
In fact, each parallel encoder requires $\log_2 M$ steps to encode the $M$-length polar codes, then $S$ steps to obtain $x$, for a total of $\log_2 M+S$ steps, i.e. slightly larger than the $\log_2 N = \log_2 M + \log_2 S$ steps required for the encoding of a standard polar code. 

\subsection{Decoding} \label{sec:dec}
\begin{algorithm}[t!]
	\caption{Sliding Window Successive Cancellation} 
	\label{alg:SWSC}
	\begin{algorithmic}[1]
		\State $\text{Load information set } \mathcal{I}$
		\State $\text{Load channel LLRs } y$
		\State $\text{Initialize buffer } l = y^{(1)}$
		\For{$s = 1 \dots S-1$}
   			\State $\hat{u}^{(s)} = \text{Decode}(l \boxplus y^{(s+1)}, \mathcal{I}_s)$ \label{alg:SC}
  			\State $\hat{x}^{(s)} = \hat{u}^{(s)} \cdot T_M$ 
 			\State $l = (-1)^{\hat{x}^{(s)}} \cdot l + y^{(s+1)}$ 
		\EndFor
   		\State $\hat{u}^{(S)} = \text{Decode}(l ,\mathcal{I}_S)$
		\State\Return $\hat{u} = [\hat{u}^{(1)},\dots,\hat{u}^{(S)}]$ 
	\end{algorithmic}
\end{algorithm}
The decoding of the proposed sliding window polar code design is performed in $S$ SC decoding steps, each one employing $M$ likelihoods, producing an estimate $\hat{u}^{(s)}$ of the partial input vector, and feeding back the partial sums as detailed in Algorithm~\ref{alg:SWSC}. 
In the following, logarithmic likelihood ratios (LLRs) are used in SC decoding\cite{balatsoukas}. 
The vector of $N$ channel LLRs is split into $S$ sub-vectors of size $M$ as $y = [y^{(1)}|y^{(2)}|\ldots|y^{(S)}]$.
The LLR buffer $l$, serving as input to each decoding step, is initialized with the first $M$ entries of $y$, $l = y^{(1)}$. 
At decoding step $s$, $s=1,2,\ldots S-1$, the decoder performs SC decoding of the $(M,K_s)$ polar code $\mathcal{P}_s$ as follows. 
The information set $\mathcal{I}_s$ is defined as in \eqref{eq:Is}, while the transformation matrix is given by $T_M$. 
The estimated input vector $\hat{u}^{(s)}$ is obtained through SC decoding of code $\mathcal{P}_s$, using vector $v=l \boxplus y^{(s+1)}$ as channel LLRs, where 
\begin{eqnarray}
A \boxplus B & = 2\tanh^{-1}(\tanh( A/2) \cdot \tanh(B/2)) \\
             &\simeq \sgn(A) \cdot \sgn(B) \cdot \min(A,B). 
\end{eqnarray}
Vector $\hat{x}^{(s)} = \hat{u}^{(s)} \cdot T_M$ is used to update the LLR buffer as 
\begin{equation*}
	l := (-1)^{x^{(s)}} \cdot l + y^{(s+1)} . 
\end{equation*}
The decoding of last partial input vector $\hat{u}^{(S)}$ is obtained using the content of the LLR buffer $l$, i.e. $v=l$. 
Finally, the input vector estimation $\hat{u}$ is given by $\hat{u} = [\hat{u}^{(1)},\dots,\hat{u}^{(S)}]$.

\begin{figure}[t!]
	\begin{center}
	    \resizebox{0.45\textwidth}{!}{\begin{tikzpicture}

\draw[thick] (-1.5,-0.5) node[left] {$y^{(s+1)}$} -- (-1,-0.5);
\draw[thick] (-1,0.5) -- (-1,-1.5);
\draw[thick] (-1,0.5) -- (2,0.5);
\draw[thick] (-1,-1.5) -- (0,-1.5);
\draw[thick] (1,-1.5) -- (6.5,-1.5);
\draw[thick] (6.5,-0.5) -- (6.5,-1.5);
\draw[->, >=stealth, thick]  (1.5,-0.5) -- (2.5,-0.5);
\draw[->, >=stealth, thick]  (2,0.5) -- (2,-0.5);
\draw[->, >=stealth, thick]  (0,-1.5) -- (0,-1);
\draw[->, >=stealth, thick]  (1,-1.5) -- (1,-1);
\draw[->, >=stealth, thick]  (6,-0.5) -- (7,-0.5) node[right] {$\hat{u}^{(s)}$};

\draw[thick] (-0.5,0) rectangle (1.5,-1) node[midway]{buffer $l$};
\draw[thick] (2.5,0) rectangle (6,-1) node[midway]{$(M,K_s)$ SC decoder};

\end{tikzpicture}}
	\end{center}
	\caption{Decoding scheme of sliding window polar codes.}
	\label{fig:dec}
\end{figure}
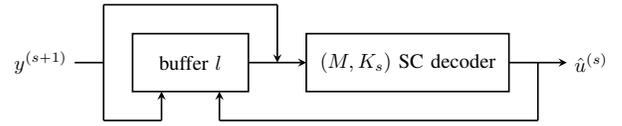

This decoding strategy can be run on-the-fly during the reception of channel outputs as depicted in Figure~\ref{fig:dec}. 
An SC list-based decoder \cite{tal_list} can be easily implemented on the basis of the described decoding algorithm.

\subsection{Performance, latency, complexity} \label{sec:perf} 

\begin{figure}[b!]
\centering
	\includegraphics[width=0.48\textwidth]{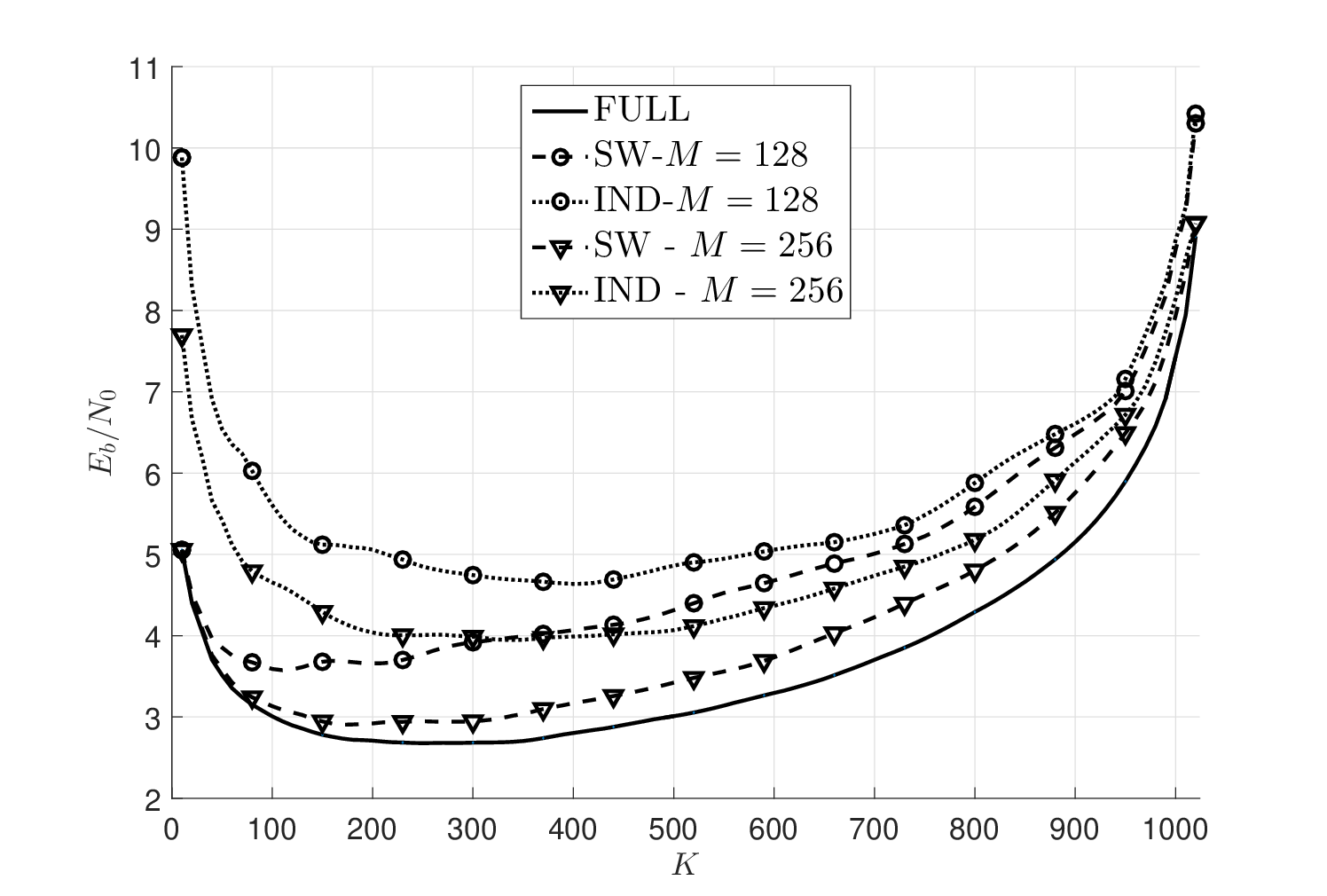}
	\caption{Minimum SNR for target BLER $10^{-3}$ with $N=1024$.}
	\label{fig:target_BLER}
\end{figure}

\begin{figure*}[t!]
	\centering
	\subfloat[$N=1024, M=128$.]{
		\includegraphics[width=0.49\textwidth]{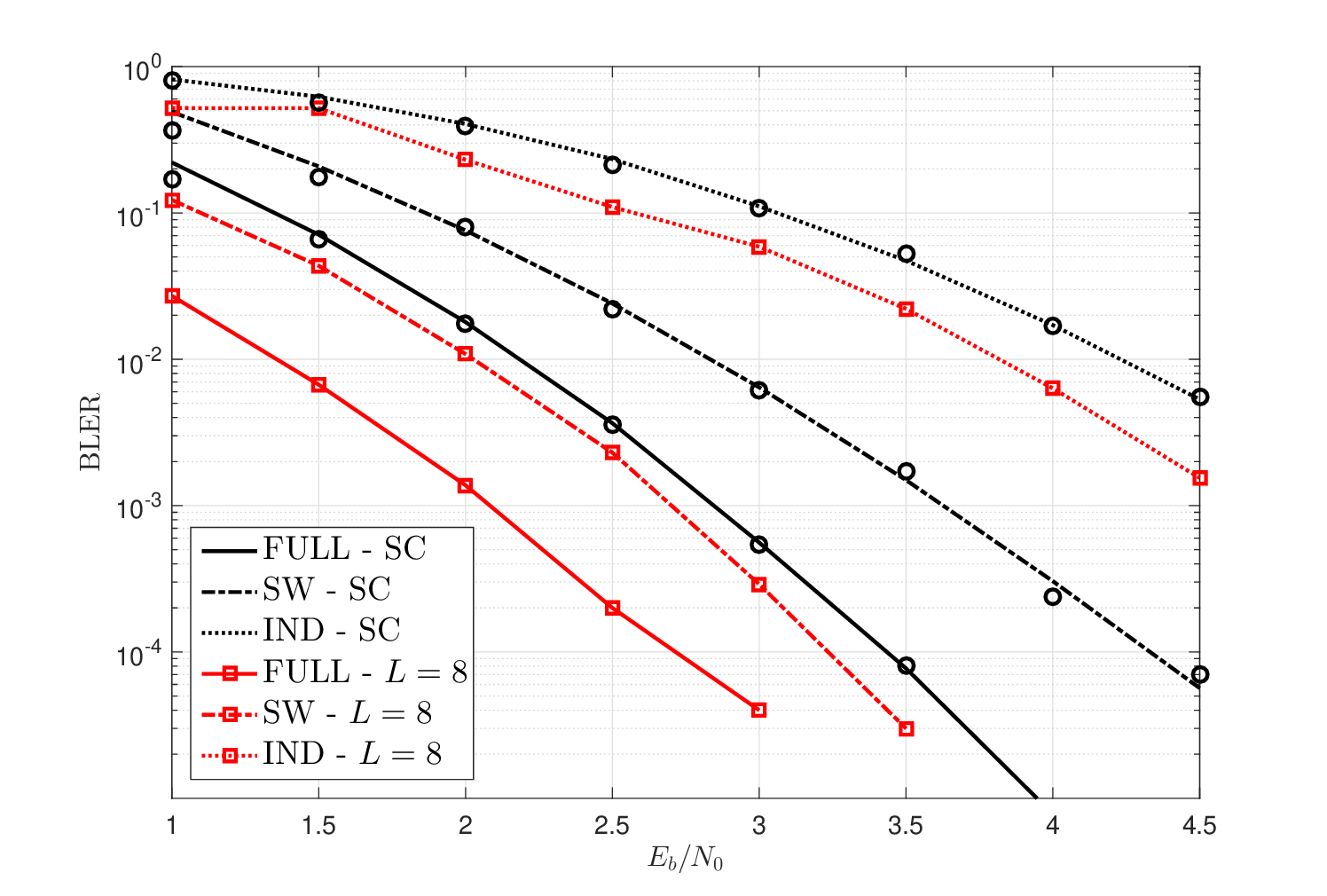}
		\label{fig:res_1024_128}
	}
	\subfloat[$N=8192, M=1024$.]{ 
		\centering
		\includegraphics[width=.49\textwidth]{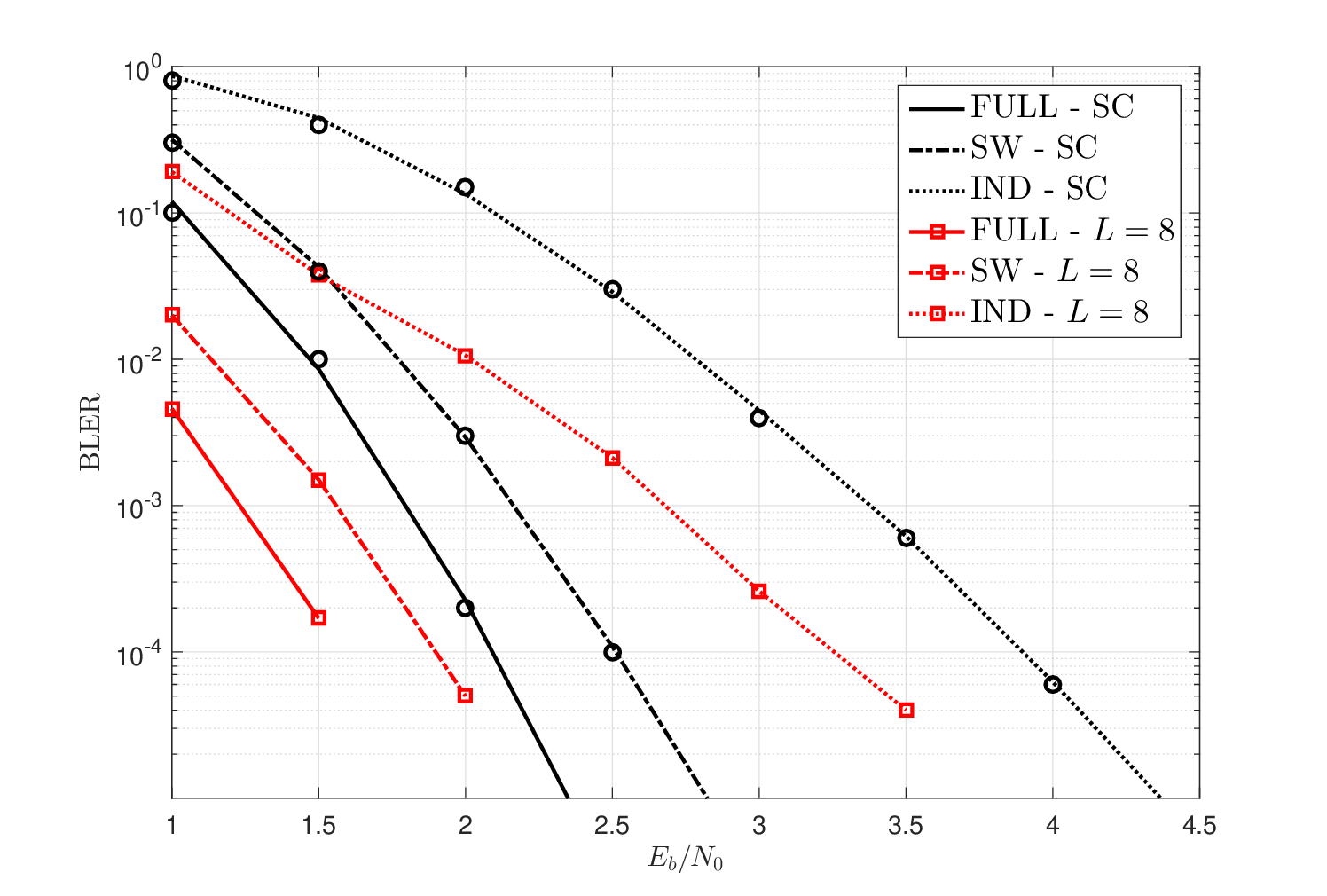}
		\label{fig:res_8192_1024}
	}
	\caption{BLER comparison of different polar codes for rate $K/N = 1/4$.}
	\label{fig:res_1_4}
\end{figure*}

Under AWGN channels it is possible to exploit the results of the DE/GA algorithm to approximate the BLER of SC decoding of polar codes. 
Considering binary phase-shift keying (BPSK) modulation, the mean values of the channel LLRs are given by $4\frac{K}{N} 10^{\gamma/10}$, where $\gamma$ represents the channel $E_b/N_0$ in dB.  
Under the Gaussian assumption, the bit error probability $P_e(u_i)$ is related to the LLR mean value $\mu_i$ as $P_e(u_i) = Q(\sqrt{\mu_i/2})$, where $Q(.)$ denotes the tail probability of the standard Gaussian distribution.
The block error probability under SC can be approximated by  
\begin{equation}
	P_e^\mathsf{SC} \sim \sum_{i \in \mathcal{I}} Q \left(\sqrt{\frac{\mu_i}{2}} \right)  .
	\label{eq:SC-BLER}
\end{equation}
Equation \eqref{eq:SC-BLER} can be used directly for sliding window polar codes designed under DE/GA. 
An estimate of the BLER under SC in case of the independent transmission of $S$ independent polar codewords with the same length $M$ is given by $P_e^\mathsf{SC} \sim 1-(1-p_e^\mathsf{SC})^S$, where $p_e^\mathsf{SC}$ is the expected BLER of a single codeword transmission, calculated according to \eqref{eq:SC-BLER}.  

Figure \ref{fig:target_BLER} shows the theoretical $E_B/N_0$ required by each code construction to reach the target BLER of $10^{-3}$ as a function of dimension $K$ for different window sizes $M$. 
Independent transmission (IND) can be seen as the state-of-the-art in the envisaged scenario: the transmitter divides the $K$ message bits into $S=N/M$ messages of $K'=K/S$ bits, that are encoded and transmitted independently using $S$ polar codes of length $M$ and dimension $K'$. 
The transmission is successful if all $S$ blocks are decoded correctly. 
In the case of full polar code (FULL) transmission, the transmitter ignores the limitations at the receiver and transmits a codeword obtained using the full $(N,K)$ polar code designed according to the best frozen set, and this is decoded at the receiver. 
Finally, in the proposed sliding window decoding (SW) the transmitter designs and encodes a polar codeword using the described technique. 
The receiver uses the proposed sliding window decoder to decode the received signal. 
We think that the proposed comparison with IND is fair, and it is the most significant given the characteristics of the proposed method, allowing to compare the performance of three codes (FULL, IND, and SW) that have different length and behavior. 
By doing so, we compare the error rate over the decoding of a block of $K$ information bits, regardless of the way they were transmitted: all in one go with a code length $N$ in case of FULL, separate in $S$ transmissions in case of IND and SW.

The proposed SW technique outperforms IND for all rates at equal window dimension, with a substantial gain of 1 dB for some of the rates. 
Moreover, for $M=256$ the sliding window polar codes show a gap of less than 0.5dB from the optimal FULL. 
Unfortunately, theoretical bounds on the BLER performance of polar codes under SCL are unknown, so it is not possible to perform a similar analysis for list decoders. 

Regarding the SC decoding latency, for the first $S-1$ phases the sliding window decoder requires $2M$ time steps to decode a length-$M$ polar code and update the buffer $l$, while $2M-2$ time steps are sufficient for the last decoding, leading to a total latency of $2MS-2=2N-2$ time steps. 
This is equal to the latency of an SC decoder for a non-systematic code of length $N$ under the same assumption. 
However, the implementation complexity of the proposed sliding window decoder is that of a polar decoder of length $M$, plus $M$ memory elements to store buffer $l$ and the logic to update it. 
We can thus conservatively identify as a complexity upper bound that of a decoder for a code of length $2M$: consequently, if $N\ge2M$, the proposed decoder always yields a lower implementation cost than a length-$N$ decoder. 
A similar analysis can be done for SCL decoding, obtaining the same outcome.

\section{Simulation results} \label{sec:simu}
In this section we present the BLER performance of the proposed sliding window design and decoding of polar codes, compared to state-of-the-art independent block transmissions and optimal full polar code transmission. 
We study a scenario where the transmitter sends $K$ bits to the receiver at a code rate $R=K/N$, but the receiver has limited decoding capabilities and can handle only $M<N$ bits.

Figure~\ref{fig:res_1_4} shows the performance of the IND, FULL and SW strategies for codes of rate $R=K/N=1/4$. 
The FULL case is used as a benchmark of the best possible achievable BLER.
The codes are decoded using either the SC or SCL algorithms with list size $L=8$. 
Black curves correspond to the SC bound calculated according to \eqref{eq:SC-BLER}, while black markers depict simulation results obtained through SC decoding. 
We can see that the bounds perfectly match the simulations, hence they can be considered as reliable approximations of the SC decoding curves for the proposed code lengths. 
Red curves with red markers correspond to SCL decoding simulations. 
The proposed solution outperforms IND under both SC and SCL decoding, while approaching the optimal performance represented by FULL.
The entity of the gain provided by SW with respect to IND is significant, corresponding to $1.5$dB. 

\section{Conclusion} \label{sec:conc}

In this work, we presented a novel multi-kernel polar code construction that allows a sliding window decoding approach, thanks to which only a fraction of the codeword bits need to be received before starting the decoding process.  
This feature will be extremely useful in future wireless 5G+ networks, where multiple devices with different computational capabilities will be served concurrently. 
The performance of the proposed construction outperforms the state-of-the-art solutions, and can approach the achievable performance with lower complexity and no cost in terms of latency. 

\bibliographystyle{IEEEbib}
\bibliography{IEEEabrv,refs}

\end{document}